\documentclass{ws-ijmpb}

\begin{document}

\markboth{Bin Liu, Ying Liang, and Shiping Feng}
{Superconductivity in Na$_{x}$CoO$_{2}\cdot y$H$_{2}$O driven by
the kinetic energy}

\catchline{}{}{}{}{}

\title{Superconductivity in Na$_{x}$CoO$_{2}\cdot y$H$_{2}$O driven by the kinetic energy}

\author{Bin Liu, Ying Liang, and Shiping Feng}
\address{Department of Physics, Beijing Normal University,
Beijing 100875, China}

%%%%%%%%%%%%%%%%%%%%%%%%%%%%%%%%%%%%%%%%%%%%%%%%%%%%%%%%%%%%
% You may repeat \author \address as often as necessary    %
%%%%%%%%%%%%%%%%%%%%%%%%%%%%%%%%%%%%%%%%%%%%%%%%%%%%%%%%%%%%

\maketitle

\begin{history}
\received{11 June 2004}
\end{history}

\begin{abstract}
Within the charge-spin separation fermion-spin theory, the
mechanism of superconductivity in Na$_{x}$CoO$_{2}\cdot y$H$_{2}$O
is studied. It is shown that dressed fermions interact occurring
directly through the kinetic energy by exchanging magnetic
excitations. This interaction leads to a net attractive force
between dressed fermions(then the electron Cooper pairs), and
their condensation reveals the superconducting ground state. The
optimal superconducting transition temperature occurs in the
electron doping concentration $\delta\approx 0.29$, and then
decreases for both underdoped and overdoped regimes, in
qualitative agreement with the experimental results.
\end{abstract}

\

The discovery of superconductivity in Na$_{0.35}$CoO$_{2}\cdot
1.3$H$_{2}$O has engendered great interest in transition metal
oxides\cite{takada}. Many experimental
measurements\cite{takada,schaak} have shown that the
superconductivity in the doped cobaltate suggests the importance
of the strong electron correlation as in the doped
cuprates\cite{schaak,tokura}, and two systems may have similar
underlying SC mechanism, i.e., it is possible that
superconductivity in the electron doped cobaltates is also driven
by the kinetic energy as in the doped cuprates\cite{feng2}. In
this paper, we will apply the charge-spin separation fermion-spin
theory\cite{feng1} to study the mechanism of superconductivity in
the doped cobaltates within the $t$-$J$ model.

The $t$-$J$ model on a triangular lattice is expressed
as\cite{baskaran,liu},
\begin{eqnarray}
H=-t\sum_{i\hat{\eta}\sigma}f_{i\sigma}^{\dagger}
f_{i+\hat{\eta}\sigma}+\mu\sum_{i\sigma}f_{i\sigma }^{\dagger}
f_{i\sigma }+J\sum_{i\hat{\eta}}{\bf S}_{i} \cdot{\bf
S}_{i+\hat{\eta}},
\end{eqnarray}
supplemented by the local constraint
$\sum_{\sigma}f^{\dagger}_{i\sigma}f_{i\sigma}\leq 1$ to remove
double occupancy, where $t<0$, $f^{\dagger}_{i\sigma}$
($f_{i\sigma}$) is the hole creation (annihilation) operator, and
${\bf S}_{i}=f^{\dagger}_{i}{\bf \sigma}f_{i}/2$ is the spin
operator in the hole representation. Then the hole operators can
be expressed\cite{feng1} as,
$f_{i\uparrow}=a^{\dagger}_{i\uparrow}S^{-}_{i}$ and
$f_{i\downarrow}=a^{\dagger}_{i\downarrow}S^{+}_{i}$, where the
spinful fermion operator $a_{i\sigma}=e^{-i\Phi_{i\sigma}}a_{i}$
describes the charge degree of freedom together with some effects
of the spin configuration rearrangements due to the presence of
the doped electron itself (dressed fermion), while the spin
operator $S_{i}$ describes the spin degree of freedom (dressed
spinon), then the single occupancy local constraint is satisfied
in analytical calculations\cite{feng1}. In this charge-spin
separation fermion-spin representation, the low-energy behavior of
the $t$-$J$ model (2) can be expressed as\cite{feng1},
\begin{eqnarray}
H&=&-t\sum_{i\hat{\eta}}(a_{i\uparrow}S^{+}_{i}
a^{\dagger}_{i+\hat{\eta}\uparrow}S^{-}_{i+\hat{\eta}}+
a_{i\downarrow}S^{-}_{i}a^{\dagger}_{i+\hat{\eta}\downarrow}
S^{+}_{i+\hat{\eta}}) \nonumber \\
&-&\mu\sum_{i\sigma}a^{\dagger}_{i\sigma} a_{i\sigma}+J_{{\rm
eff}}\sum_{i\hat{\eta}}{\bf S}_{i}\cdot {\bf S}_{i+\hat{\eta}},
\end{eqnarray}
with $J_{{\rm eff}}=(1-\delta)^{2}J$, and $\delta=\langle
a^{\dagger}_{i\sigma}a_{i\sigma}\rangle=\langle a^{\dagger}_{i}
a_{i}\rangle$ is the electron doping concentration. In this case,
the magnetic energy ($J$) term in the $t$-$J$ model is only to
form an adequate dressed spinon configuration, while the kinetic
energy ($t$) term has been transferred as the dressed
fermion-spinon interaction, which dominates the essential physics.
This dressed fermion-spinon interaction is quite strong, and can
induce the dressed fermion pairing state (then the electron
pairing state and superconductivity) by exchanging dressed spinon
excitations in a higher power of the electron doping concentration
$\delta$. As in the conventional superconductors, the order
parameter for the electron Cooper pair can be expressed as
$\Delta=\langle f^{\dagger}_{i\uparrow}f^{\dagger}_{j\downarrow}-
f^{\dagger}_{i\downarrow}f^{\dagger}_{j\uparrow}\rangle =\langle
a_{i\uparrow}a_{j\downarrow}S^{+}_{i}S^{-}_{j}-a_{i\downarrow}
a_{j\uparrow}S^{-}_{i}S^{+}_{j}\rangle$.  In the doped regime
without AFLRO, the dressed spinon correlation function $\langle
S^{+}_{i}S^{-}_{j}\rangle=\langle S^{-}_{i} S^{+}_{j} \rangle$,
and the order parameter for the electron Cooper pair can be
written as $\Delta=-\langle S^{+}_{i}S^{-}_{j} \rangle\Delta_{a}$,
with the dressed fermion pairing order parameter
$\Delta_{a}=\langle a_{j\downarrow} a_{i\uparrow}-
a_{j\uparrow}a_{i\downarrow}\rangle$. Following the discussions in
Ref. 4, we can obtain the dressed fermion pair order parameter as,
\begin{eqnarray}
\Delta^{(d)}_{a}={2\over N}\sum_{k}\mid\gamma^{(d)}_{k}\mid^{2}
{\bar{\Delta}^{(d)}_{a}\over E_{k}}{\rm th} [{1\over 2}\beta E_{
k}].
\end{eqnarray}
where $\gamma^{(d)}_{k}=d_{1k}+id_{2k}$, $d_{1k}= 2{\rm
cos}k_{x}-{\rm cos}[(k_{x}- \sqrt{3} k_{y})/2]-{\rm cos}[(k_{x}+
\sqrt{3}k_{y})/2]$ and $d_{2k}=\sqrt{3}{\rm cos}
[(k_{x}+\sqrt{3}k_{y})/2] -\sqrt{3}{\rm cos} [(k_{x}-
\sqrt{3}k_{y})/2]$, and other physical quantities have been given
in Ref. 4. For a discussion of the physical properties of the SC
state, we now need to calculate the electron off-diagonal Green's
function $\Gamma^{\dagger}(i-j,t-t')=\langle\langle
f^{\dagger}_{i\uparrow}(t);f^{\dagger}_{j\downarrow}(t')\rangle
\rangle$. With the help of electron off-diagonal Green's function
and in the framework of the charge-spin separation fermion-spin
theory\cite{feng1}, we obtain the SC gap function as,
\begin{eqnarray}
\Delta^{(d)}(k)&=&-{1\over N}\sum_{p}
{\bar{\Delta}^{(d)}_{a}(p-k)B_{p}\over 4E_{p-k}\omega_{p}}{\rm th}
{\beta E_{p-k}\over 2} {\rm coth}{\beta\omega_{p}\over 2},
\end{eqnarray}
which shows that the symmetry of the electron Cooper pair is the
same as the symmetry of the dressed fermion pair, i.e., the SC gap
function can be written as $\Delta^{(d)}(k)=\Delta^{(d)}
(d_{1k}+id_{2k})$. The SC gap function in Eq. (4) indicates that
the SC transition temperature $T_{c}$ occurring in the case of
$\Delta^{(d)}=0$ is identical to the dressed fermion pair
transition temperature occurring in the case of
$\bar{\Delta}^{(d)}_{a}=0$. This SC transition temperature $T_{c}$
as a function of electron doping concentration $\delta$ in the
d-wave symmetry for $t/J=-2.5$ is plotted in Fig. 1 in comparison
with the experimental data\cite{schaak}(inset). It is shown that
the maximal SC transition temperature T$_{c}$ occurs around the
optimal electron doping concentration $\delta\approx 0.29$, and
then decreases for both underdoped and overdoped regimes. Using an
reasonable estimation value of $J\sim 10$mev to 20mev in
Na$_{x}$CoO$_{2}\cdot y$H$_{2}$O, the SC transition temperature in
the optimal doping is T$^{{\rm optimal}}_{c}\approx 0.02J\approx
3{\rm K}\sim 6{\rm K}$. Our these results are in qualitative
\begin{figure}[th]
\centerline{\psfig{file=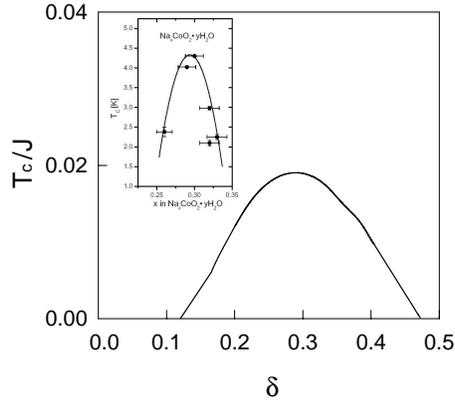,width=6cm}} \caption{The
superconducting transition temperature as a function of the
electron doping concentration in the d-wave symmetry for
$t/J=-2.5$. Inset: the experimental result on
Na$_{x}$CoO$_{2}\cdot y$H$_{2}$O taken from Ref. 2.}
\end{figure}
agreement with the experimental data\cite{schaak}. Since
Na$_{x}$CoO$_{2}\cdot y$H$_{2}$O is the electron doped Mott
insulator on a triangular lattice, therefore the system has strong
geometrical spin frustration. This magnetic frustration also
induces the strong charged carrier's quantum fluctuation. Both
strong magnetic frustration and charged carrier's quantum
fluctuation suppress the SC transition temperature.
%%%%%%%%%%%%%%%%%%%%%%%%%%%%%%%%%%%%%%%%%%%%%%%%%%%%%%%%%%%%
% Doing Acknowledgement                          %
%%%%%%%%%%%%%%%%%%%%%%%%%%%%%%%%%%%%%%%%%%%%%%%%%%%%%%%%%%%%

\section*{Acknowledgements}

This work was supported by the National Natural Science Foundation
of China under Grant Nos. 10125415, 10074007, and 10347102, and
the Grant from Beijing Normal University.

%%%%%%%%%%%%%%%%%%%%%%%%%%%%%%%%%%%%%%%%%%%%%%%%%%%%%%%%%%%%
% Doing Appendix(ices) - Appendix A & B are shown below    %
%%%%%%%%%%%%%%%%%%%%%%%%%%%%%%%%%%%%%%%%%%%%%%%%%%%%%%%%%%%%

%%%%%%%%%%%%%%%%%%%%%%%%%%%%%%%%%%%%%%%%%%%%%%%%%%%%%%%%%%%%
% Doing references:                                %
%%%%%%%%%%%%%%%%%%%%%%%%%%%%%%%%%%%%%%%%%%%%%%%%%%%%%%%%%%%%

\end{document}